\documentclass[aps,superscriptaddress,prd,
onecolumn,
floatfix, 
nofootinbib,
amsmath,amssymb,amsfonts,longbibliography]{revtex4-2}

\usepackage{mathrsfs,amsmath}
\usepackage{mathtools}
\usepackage{empheq}
\usepackage[pdftex]{graphicx}
\usepackage[dvipsnames]{xcolor}
\usepackage{float}
\usepackage{physics}
\usepackage{comment}
\usepackage[normalem]{ulem}
\usepackage{caption}
\usepackage{subcaption}
\usepackage[colorlinks,linkcolor=blue,anchorcolor=violet,citecolor=red]{hyperref}

\usepackage{bm}

\newcommand*{\Blue}{\textcolor[rgb]{0.0,0.0,0.8}}

\newcommand*{\Orange}{\textcolor[rgb]{0.8,0.4,0.1}}

\begin{document}

\title{
Boosting Gravity-Induced Entanglement through Parametric Resonance
}

\author{Yuka Shiomatsu}
\email{g2540616@edu.cc.ocha.ac.jp}
\affiliation{Department of Physics, Ochanomizu University, Bunkyo, Tokyo 112-8610, Japan}
\author{Youka Kaku}
\email{youkakaku@pegasus.kobe-u.ac.jp}
\affiliation{Department of Physics, Kobe University, Kobe 657-8501, Japan}
\author{Akira Matsumura}
\email{matsumura.akira@phys.kyushu-u.ac.jp}
\affiliation{Department of Physics, Kyushu University, Fukuoka 819-0395, Japan}
\affiliation{Quantum and Spacetime Research Institute, Kyushu University, Fukuoka 819-0395, Japan}
\author{Tomohiro Fujita}
\email{fujita.tomohiro@ocha.ac.jp}
\affiliation{Department of Physics, Ochanomizu University, Bunkyo, Tokyo 112-8610, Japan}
\affiliation{Kavli Institute for the Physics and Mathematics of the Universe (Kavli IPMU),
WPI, UTIAS, The University of Tokyo, Kashiwa, Chiba 277-8568, Japan}

\begin{abstract}
Establishing quantum gravity theory remains one of the major challenges in modern physics, as the lack of experimental evidence makes it difficult to explore. 
In response to this challenge, proposals to test quantum entanglement induced by Newtonian gravity in table-top experiments have attracted significant attention as a potentially feasible approach far below the Planck energy scale.
In this work, we propose a scheme to amplify gravity-induced entanglement between two masses using parametric resonance. Specifically, we consider two parametrically resonant oscillators interacting through Newtonian gravity, each governed by the Mathieu equation.
We analyzed the logarithmic negativity between two oscillators and investigate the effects of random force noise and linear damping.
As a result, we find an exponential growth of gravity-induced entanglement between the oscillators, which reflects the dynamical instability of parametric resonant systems.
\end{abstract}

\maketitle




\section{Introduction}

Combining quantum theory and gravity has been one of the major challenges in modern physics. One of the main reasons for this difficulty is the lack of experimental evidence to test the quantum aspects of gravity. 
Recently, several works~\cite{Bose2017,Marletto2017} have proposed testing the quantum nature of Newtonian gravity in table-top experiments as a first step toward revealing quantum gravity. In these proposals, two masses interacting through Newtonian gravity are considered, and they aim to clarify whether such an interaction can generate quantum entanglement between the masses. 
Their argument builds on a fundamental concept of quantum information theory, namely that local operation and classical communication (LOCC) cannot generate quantum entanglement. Therefore, if two masses interacting exclusively through Newtonian gravity become entangled from an initially separable state, it follows that the time evolution under gravitational interaction cannot be characterized as LOCC.
These ideas have attracted considerable attention because they aim to probe quantum aspects of gravity at energy scales far below the Planck scale, making them much more feasible than earlier proposals such as the direct detection of gravitons~\cite{Dyson2013}. Their proposals inspired a wide range of related studies using optomechanical systems~\cite{Balushi2018,Matsumura2020,Miao2020,Miki2024,Miki2024f,Miki2025}, mechanical oscillators~\cite{Qvarfort2020,Kaku2022}, matter-wave interferometor~\cite{Van2020,Chevalier2020,Nguyen2020,Torovs2021,Miki2021}, and their hybrid systems~\cite{Carney2021,carney2022,Matsumura2021,Streltsov2022}.

Despite the growing interest, the experimental realization of gravity-induced entanglement still remains highly challenging~\cite{Rijavec2021}. The primary obstacle is that gravitational interactions are extremely weak, making it difficult to generate a detectable amount of entanglement. In realistic setups, environmental decoherence caused by air molecules, blackbody radiation, and various external forces controlling the probe system further suppresses entanglement generation. 
While microscopic systems, such as the latest quantum control on an acoustic-wave resonator with a mass of $1.6\times 10^{-5}\,\rm g$~\cite{Bild2023}, are favorable for minimizing environmental noise, macroscopic systems, such as the $9\times 10^{-2}\, \rm g$ mass used in the recent precise measurements of the gravitational constant~\cite{Westphal2021}, are required to enhance gravitational coupling. This fundamental trade-off prevents straightforward realization of these proposals. 
Several theoretical works have explored strategies to amplify gravity-induced entanglement and overcome this limitation~\cite{Krisnanda2020,Weiss2021,Pedernales2022,Kaku2024,Braccini2024,Kaku2025}.
For instance, in Refs. \cite{Krisnanda2020,Fujita2023}, the probe is assumed to be released or to follow an unstable potential, leading to a rapid growth of gravity-induced entanglement due to position spreading. Similarly, the authors of Refs.~\cite{Weiss2021,Braccini2024} consider periodic switching of the probe???s effective potential between harmonic and inverted-harmonic oscillator configurations, which again results in exponential enhancement of the gravity-induced entanglement via position spreading.

In this paper, we address this issue by proposing an improved setup that enhances gravity-induced entanglement using dynamical instability. In particular, we utilize parametric resonance, a phenomenon in which the output amplitude grows exponentially due to periodic modulation of the system parameters, analogous to a swing being pushed at the right timing.
Specifically, we consider two parametric resonant oscillators, each governed by the Mathieu equation, and coupled via first-quantized Newtonian gravity. We then evaluate the logarithmic negativity as a measure of the entanglement between the two systems. We find that the entanglement increases dramatically in the unstable parameter region where parametric resonance occurs. Moreover, when the resonance dominates, the entanglement grows exponentially with a rate equal to the characteristic exponent of the Mathieu equation. 
In contrast to previous studies \cite{Weiss2021,Braccini2024,Kaku2024}, we demonstrate that an exponential growth of gravity-induced entanglement can be achieved even within a stable trapping potential, simply by modulating its curvature periodically, without introducing any unstable potential. 
Such periodic modulation has been experimentally demonstrated in Ref.~\cite{Hartwig2023}, although their purpose was to stabilize the probe system rather than to drive its dynamical instability. A similar setup could be realized by periodically modulating the radiation pressure of an optical spring acting on a mechanical system~\cite{Rossi2025,Ma2020}.
Finally, we analyze the decoherence effects of stochastic forces and dissipation.

The remainder of this paper is organized as follows.
In Section \ref{sec:Model}, we briefly review the concept of parametric resonance in a single oscillator described by the Mathieu equation. We then formulate the time evolution of two coupled parametric oscillators interacting via Newtonian gravity and evaluate their logarithmic negativity. Section \ref{sec:GIE} presents numerical results and analytical approximations for the gravity-induced entanglement. In Section \ref{sec:damping_noise}, we extend the analysis to include stochastic noise and dissipation from the environmental system. Finally, Section \ref{sec:conclusion} summarizes our findings and provides concluding remarks.

\section{\Blue{Analytical Model and Entanglement Dynamics without Environment}}\label{sec:Model}

\begin{figure}[htbp]
    \centering
    \includegraphics[width=0.7\linewidth]{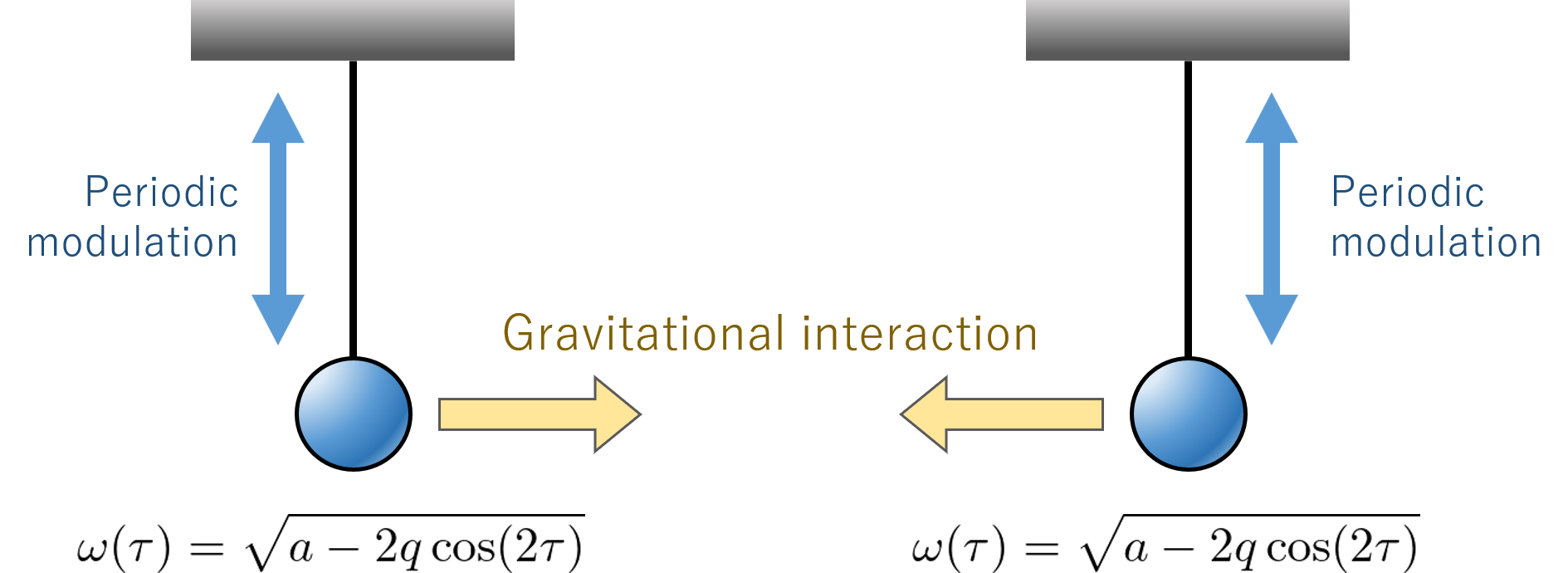}
    \caption{Setup with two parametric resonant oscillators coupled via Newtonian gravity.
    Each oscillator is assumed to follow the Mathieu equation, which describes a periodical modulation of its effective frequency as $\omega(\tau)=\sqrt{a-2q \cos(2\tau)}$.}

    \label{fig:setup}
\end{figure}

In this work, we consider a model of two Mathieu oscillators coupled via Newtonian gravity, as depicted in Fig.~\ref{fig:setup}.

The model builds upon the canonical parametric resonant oscillator, which is described by the Mathieu equation~\cite{Landau1976}:
\begin{equation}
  \frac{d^{2}X}{d\tau^{2}}+\bigl[a-2q\cos(2\tau)\bigr]X=0\,,
  \label{eq:single-mathieu}
\end{equation}
where $\tau$ is a dimensionless time. Two model parameters, $a$ and $q$, 
govern the stability of the solutions. Depending on their values, the system exhibits either stable oscillations or exponential growth of displacement, with the growth rate depending sensitively on $(a,q)$.
We focus on the region $a \geq 2q$, where the effective squared frequency remains non-negative at all times. 

\begin{figure}[htbp]
  \centering
  \includegraphics[width=0.9\linewidth]{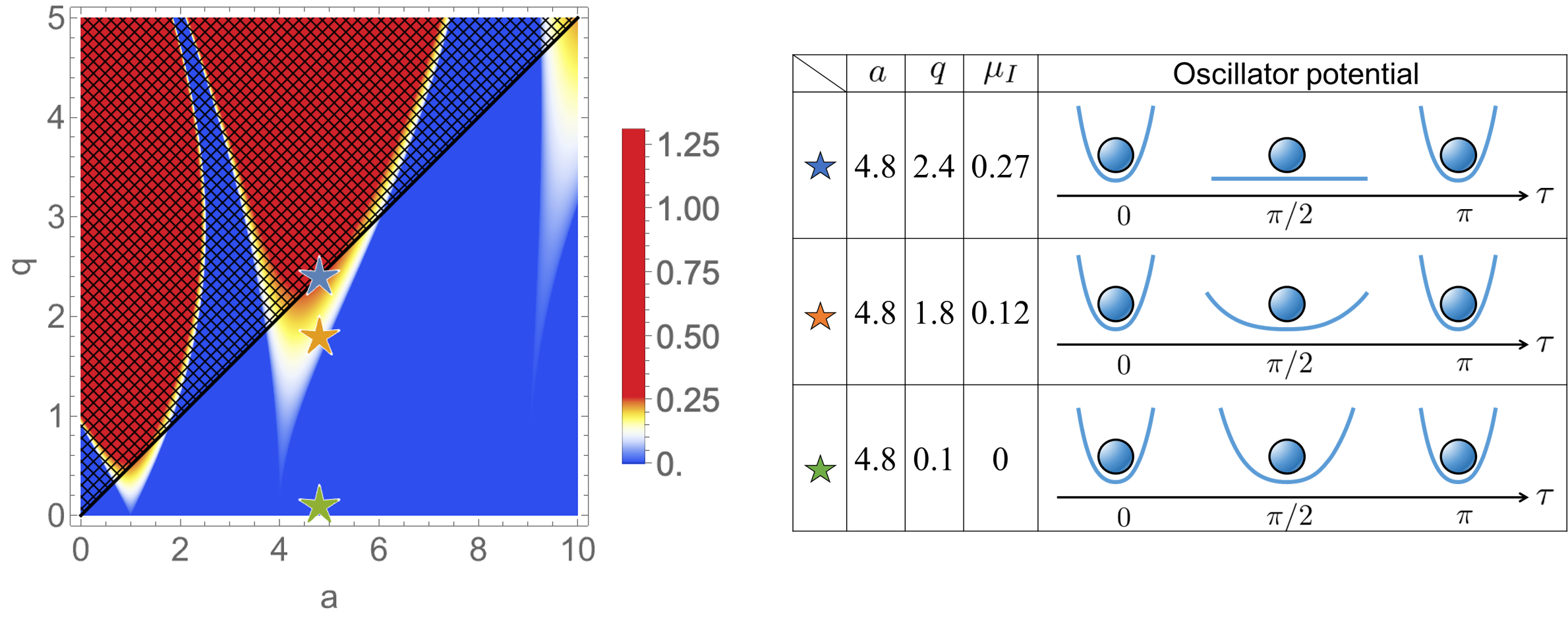}
  \caption{
  Left panel: Stability diagram of the Mathieu equation~\eqref{eq:single-mathieu} in the \((a,q)\)-plane. The instability rate, defined as \(\mu_I\equiv -{\rm Im}(\mu)\), is represented by a color scale: blue regions $(\mu_I=0)$ correspond to stable solutions , while red regions $(\mu_I>0.25)$ indicate strong exponential growth. The upper-left region with \(2q>a\), where \(\omega^2\) becomes negative during part of the cycle, is cross-hatched in black; such cases are challenging to realize experimentally and are not considered in this work (see however \cite{Fujita2023,Michimura2017}). 
  Right panel: Periodic modulation of the oscillator potential for representative parameter points shown in the left panel. The blue, orange and green markers correspond to $(a, q) = (4.8,2.4)$, $(4.8,1.8)$, and $(4.8,0.1)$, respectively. As $q$ increases, the potential becomes shallower at $\tau=\pi(n+\frac{1}{2})$ ($n$ is an integer).
  }
  \label{fig:strutt}
\end{figure}
The left panel of Fig.~\ref{fig:strutt} shows the stability diagram in the \((a,q)\)-plane.
Because the effective squared frequency 
\(\omega^2(\tau)=a-2q\cos(2\tau)\) is periodic in \(\tau\), 
the solutions to Eq.~\eqref{eq:single-mathieu} take the Floquet form $   X(\tau)=\exp(i\mu\tau)f(\tau)$ where $f(\tau)$ is a periodic function of order unity $f(\tau+\pi)=f(\tau)$.
If the Floquet exponent $\mu$ is purely real, 
the amplitude remains bounded (stable), whereas a nonzero negative imaginary part 
leads to exponential growth of $X(\tau)$ (unstable). 
In the left panel of Fig.~\ref{fig:strutt}, the negative of the imaginary part of the Floquet exponent $\mu_I\equiv -{\rm Im}(\mu)$ is represented by color scale.
In addition, we highlight three representative parameters points. The blue, orange and green markers corresponds to $(a, q,\mu_I) = (4.8,2.4,0.27)$, $(4.8,1.8,0.12)$, and $(4.8,0.1,0)$, respectively. These parameters will be used later in the analysis presented in Section \ref{sec:GIE}.
In the right panel, periodic modulation of the oscillator potential for these representative parameters are shown.
As $q$ increases, the potential becomes shallower at $\tau=\pi(n+\frac{1}{2})$ ($n$ is an integer), resulting in a stronger spatial broadening of the wavefunction. 
How much each oscillator???s wavefunction spreads essentially determines the amount of gravity-induced entanglement: the larger the spreading, the stronger the entanglement.
Since wavefunctions of Mathieu oscillators grow exponentially in the unstable regime, we use this setting for the gravitationally coupled pair.

In the rest of this section, we formulate the model and provide an analytical treatment of the entanglement dynamics without environmental effects. 
We consider the following Hamiltonian for two parametric resonant oscillators coupled by gravity:

\begin{equation}
  \hat{H}(t)=\frac{\hbar\omega}{2}\sum_{i=1}^{2}\!\Bigl[\hat{P}_i^{2}+\bigl(a-2q\cos(2\tau)\bigr)\hat{X}_i^{2}\Bigr]
  \;+\;\hbar\omega\,\eta\,\hat{X}_{1} \hat{X}_{2}\,,
  \label{eq:model-H}
\end{equation}
where we introduce the dimensionless canonical variables
\begin{equation}
  \hat{X}_i\equiv \sqrt{\frac{m_i\omega}{\hbar}}\,\hat{x}_i,\qquad
  \hat{P}_i\equiv \frac{\hat{p}_i}{\sqrt{\hbar m_i\omega}},\qquad
\end{equation}
satisfying the usual commutation relation $[\hat{X}_i,\hat{P}_j]=\mathrm{i}\,\delta_{ij}$.
We also define the dimensionless time \(\tau=\omega t\) and the standard Mathieu parameters $a$ and $q$.
Note that in this paper, we focus on the region  $a\geq 2q$, where the effective squared frequency remains non-negative at all times.
The last term in \eqref{eq:model-H} represents the gravitational interaction. The Newtonian potential between the two oscillators can be expanded for the small relative displacement $|\hat{x}_1-\hat{x}_2|\ll d$ as
\begin{equation}
  V_{\mathrm{grav}}=-\frac{Gm_1m_2}{|\,d+\hat{x}_1-\hat{x}_2\,\bigr|}
  \simeq -\frac{Gm_1m_2}{d}
  + \frac{Gm_1m_2}{d^{2}}(\hat{x}_1-\hat{x}_2)
  - \frac{Gm_1m_2}{d^{3}}(\hat{x}_1-\hat{x}_2)^2\,.
  \label{Vgrav}
\end{equation}
Here, the leading interaction is the cross term in the last term, and it can be rewritten as $\hbar \omega \eta \hat{X}_1 \hat{X}_2$ with the coefficient
\begin{equation}
 \eta \equiv \frac{2G\sqrt{m_{1}m_{2}}}{d^{3}\omega^{2}}
 \approx 10^{-12}\,
    \left(\frac{\omega}{\rm 1~kHz}\right)^{-2}
    \left(\frac{\sqrt{m_1 m_2}/d^3}{\rm 8~g/cm^3}\right)\,.
\label{eta}
\end{equation}
This fiducial frequency $1$kHz is adopted as an example~\cite{Fujita2023}.
The smallness of $\eta$ reflects the weakness of gravity, which poses a challenge we address in what follows.
We ignore the other potential terms for simplicity  in Eq.~\eqref{Vgrav}. 

We work in the Heisenberg picture. From the Hamiltonian~\eqref{eq:model-H}, we obtain the coupled second-order differential equations as
\begin{align}
  \frac{d^{2}\hat{X}_{1}}{d\tau^{2}}+\big(a-2q\cos(2\tau)\big)\hat{X}_{1}+\eta\,\hat{X}_{2}&=0,
  \notag
  \\
  \frac{d^{2}\hat{X}_{2}}{d\tau^{2}}+\big(a-2q\cos(2\tau)\big)\hat{X}_{2}+\eta\,\hat{X}_{1}&=0.
  \label{eq:coupled-second}
\end{align}
We introduce the linearly combined variables
\begin{equation}
  \hat{X}_{\pm} \equiv \frac{\hat{X}_{1}\pm \hat{X}_{2}}{\sqrt{2}},\qquad
  \hat{P}_{\pm} \equiv \frac{\hat{P}_{1}\pm \hat{P}_{2}}{\sqrt{2}}\,.
  \label{pm variable}
\end{equation}
With these new variables, the above coupled equations decouple into two independent differential equations,
\begin{equation}
  \frac{d^{2}\hat{X}_{\pm}}{d\tau^{2}}+\Bigl[\bigl(a\pm\eta\bigr)-2q\cos(2\tau)\Bigr]\hat{X}_{\pm}=0,
  \label{eq:mathieu-pm}
\end{equation}
which take the form of well-known Mathieu equations with shifted parameters $a\pm \eta$.

We denote the even/odd Mathieu solutions by \(C\) and \(S\).
Writing \(C_\pm(\tau)\equiv C(a\pm\eta, q, \tau)\) and \(S_\pm(\tau)\equiv S(a\pm\eta, q, \tau)\), the general solutions to Eqs.~\eqref{eq:mathieu-pm} are
\begin{equation}
  \hat{X}_{\pm}(\tau)=\hat{A}_{\pm}\,C_\pm(\tau)+\hat{B}_{\pm}\,S_\pm(\tau),\qquad
  \hat{P}_{\pm}(\tau)=\frac{d\hat{X}_{\pm}}{d\tau}
  =\hat{A}_{\pm}\frac{d C_\pm}{d\tau}(\tau)+\hat{B}_{\pm}\frac{d S_\pm}{d\tau}(\tau),
  \label{eq:op-solution}
\end{equation}
where $\hat{A}_\pm, \hat{B}_\pm$ are operator coefficients which will be fixed in the following.
For the covariance analysis, we assume a centered Gaussian state, i.e. 
\(\langle \hat X_{1,2}\rangle=\langle \hat P_{1,2}\rangle=0\) at \(\tau=0\).
Because the dynamics are linear and homogeneous, the first moments remain zero for all \(\tau\).
Taking the expectation value of $\hat{X}_\pm (\tau)$ given in Eqs.~(9) then leads to 
\(\langle \hat X_{\pm}(\tau)\rangle=\langle \hat A_{\pm}\rangle C_{\pm}(\tau)+\langle \hat B_{\pm}\rangle S_{\pm}(\tau)\);
since \(C_{\pm}(\tau)\) and \(S_{\pm}(\tau)\) have distinct time dependencies,
this implies
\(\langle \hat A_{\pm}\rangle=\langle \hat B_{\pm}\rangle=0\)\,.

The commutation relation requires
\begin{equation}
  [\hat{A}_\lambda,\hat{B}_\lambda]
  =\frac{\mathrm{i}}{\,C_\lambda(0)\,\tfrac{d S_\lambda}{d\tau}(0)\,},\qquad
  [\hat{A}_\lambda,\hat{A}_\sigma]
  =[\hat{B}_\lambda,\hat{B}_\sigma]
  =[\hat{A}_\lambda,\hat{B}_\sigma]=0\ \quad(\lambda \neq \sigma),
  \label{eq:AB-comm}
\end{equation}
with $\lambda, \sigma=\pm$. Here, the first condition follows from the commutation relation $[\hat{X}_\lambda(\tau),\hat{P}_\sigma(\tau)]=\mathrm{i}\delta_{\lambda\sigma}$.
To evaluate it, recall that the Wronskian of the Mathieu solutions is constant,
\begin{equation}
W_\lambda(\tau) \equiv C_\lambda(\tau)\tfrac{d S_\lambda}{d\tau}(\tau)-S_\lambda(\tau)\tfrac{d C_\lambda}{d\tau}(\tau)={\rm const.}    
\end{equation}
Since the even and odd solutions satisfy \(S_\lambda(0)=0\) and \(\tfrac{d C_\lambda}{d\tau}(0)=0\), the above constant reduces to $W_\lambda(0)=C_\lambda(0)\tfrac{d S_\lambda}{d\tau}(0)$ at $\tau=0$.

Now let us consider the covariance matrix $V(\tau)$ defined by
\begin{equation}
V_{ij}(\tau)=\frac{1}{2}\left\langle \hat{R}_i(\tau)\hat{R}_j(\tau)+\hat{R}_j(\tau)\hat{R}_i(\tau)\right\rangle\,,\qquad
\hat{\mathbf{R}}(\tau)=(\hat{X}_1,\hat{P}_1,\hat{X}_2,\hat{P}_2)^\top\,.
\end{equation}
We assume that the initial state is the separable Gaussian ground state with the covariance $V(0)=\tfrac12 I_4$ at \(\tau=0\), which determines the correlations of the operator coefficients:
\begin{equation}
  \langle \hat{A}_\lambda^{\,2}\rangle=\frac{1}{2\,C_\lambda(0)^2},\qquad
  \langle \hat{B}_\lambda^{\,2}\rangle=\frac{1}{2\,\bigl(\tfrac{d S_\lambda}{d\tau}(0)\bigr)^{\!2}},\qquad
  \langle \hat{A}_\lambda\hat{B}_\sigma\rangle
  =-\langle \hat{B}_\lambda\hat{A}_\sigma\rangle
  =\frac{\mathrm{i}}{2\,C_\lambda(0)\,\tfrac{d S_\lambda}{d\tau}(0)}\,\delta_{\lambda\sigma}.
  \label{eq:AB-moments}
\end{equation}
\begin{equation}
\langle \hat A_\pm\hat A_\mp\rangle=\langle \hat B_\pm\hat B_\mp\rangle=0.
\label{14}
\end{equation}
The commutation relations~\eqref{eq:AB-comm} together with the initial Gaussian state determine all correlations among the operator coefficients, which in turn completely specifies the covariance $V(\tau)$.
The obtained covariance matrix can be written in the compact form
\begin{equation}
  V(\tau)=\frac{1}{4}\sum_{\lambda=\pm}
  \begin{pmatrix}
    v_\lambda(\tau) & \lambda\,v_\lambda(\tau) \\
    \lambda\,v_\lambda(\tau) & v_\lambda(\tau)
  \end{pmatrix}\,,
  \label{eq:V-compact}
\end{equation}
where we define
\begin{equation}
      \mathbf{C}_\lambda(\tau)=
    \begin{pmatrix} C_\lambda(\tau)/C_\lambda(0)\\[2pt] \tfrac{d C_\lambda}{d\tau}(\tau)/C_\lambda(0)\end{pmatrix},\qquad
  \mathbf{S}_\lambda(\tau)=
    \begin{pmatrix} S_\lambda(\tau)/\tfrac{d S_\lambda}{d\tau}(0)\\[2pt] \tfrac{d S_\lambda}{d\tau}(\tau)/\tfrac{d S_\lambda}{d\tau}(0)\end{pmatrix},
\end{equation}
and the $2\times 2$ matrix \(v_\lambda(\tau)=\mathbf{C}_\lambda(\tau)\mathbf{C}^{\!\top}_\lambda(\tau)+\mathbf{S}_\lambda(\tau)\mathbf{S}^{\!\top}_\lambda(\tau)\).

To quantify the entanglement generated between the two oscillators by gravity, we use the logarithmic negativity \(E_{\mathcal N}\),
\begin{equation}
  E_{\mathcal N}(\tau)\equiv \log_{2}\!\bigl(2\,\mathcal N(\tau)+1\bigr).
  \label{eq:LogNegativity}
\end{equation}
Here, the negativity $\mathcal N$ \cite{Vidal2002} can be computed from the covariance matrix via the standard formula~\cite{Weedbrook2012}
\begin{equation}
  \mathcal N(\tau)
  = \max\!\Biggl\{0,\;
     \frac{1}{2}\!\biggl(
       \frac{1}{\sqrt{2}}\,
       \Bigl[\tilde\Delta-\sqrt{\tilde\Delta^{2}-4\det V}\Bigr]^{-1/2}
       -1\biggr)\Biggr\},
\end{equation}
where the covariance matrix $V$ is expressed in the block form as
\begin{equation}
  V=\begin{pmatrix} A & C\\ C^{\mathsf T} & B \end{pmatrix},\qquad
  \tilde\Delta\equiv \det A+\det B-2\det C .
\end{equation}
Note that in the absence of damping or decoherence effects, the evolution is symplectic, hence the determinant is conserved, \(\det V(\tau)=\det V(0)=1/16\). This invariance will be broken once damping or decoherence is introduced, as we will discuss in Sec.~\ref{sec:damping_noise}.

\section{Entanglement Results without Environment}
\label{sec:GIE}

In this section, we investigate the generation of gravity-induced entanglement based on the analytic solutions derived in the previous section.

\begin{figure}[t]
\centering
\includegraphics[width=.6\linewidth]{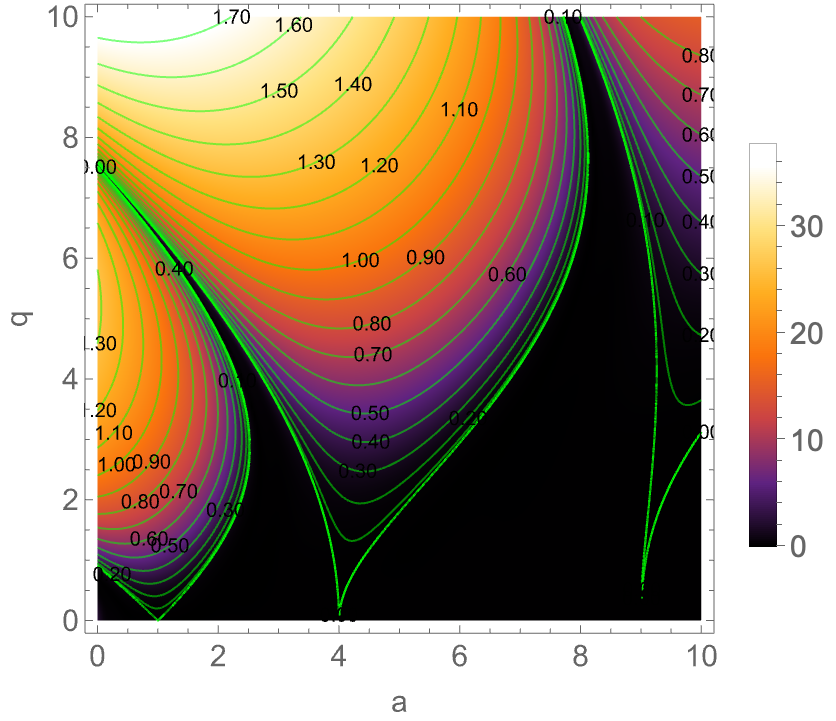}
\caption{Logarithmic negativity $E_{\mathcal N}$ evaluated at $\tau=3\pi$ for a relatively large gravitational coupling $\eta=0.01$ in the $(a,q)$-plane.
The color map, ranging from dark purple to bright yellow, shows the magnitude of the generated entanglement $E_{\mathcal N}$.
Thin green lines represent contours of the imaginary part of the Floquet exponent $\mu_I$ at intervals of 0.1, and the thick green curves trace $\mu_I=0$, marking the boundary between stable and unstable regions. As expected, $E_{\mathcal N}$ increases with the strength of the Mathieu instability $\mu_I$.}
  \label{fig:ENmap}
\end{figure}
Fig.\,\ref{fig:ENmap} shows the distribution of entanglement on the \((a,q)\) plane. The horizontal axis is \(a\), the vertical axis is \(q\), and the color map corresponds to the generated logarithmic negativity \(E_{\mathcal N}\). Brighter regions indicate larger \(E_{\mathcal N}\), i.e., stronger entanglement. 
The green contours indicate the imaginary part of the Floquet exponent, \(\mu_I\).
As expected, Fig.\,\ref{fig:ENmap} shows that gravitational entanglement remains negligible in the stable region outside the thick green curves, whereas strong entanglement emerges where $\mu_I$ is large. Hence, parameter choices with stronger instability are advantageous for entanglement generation.
Note that, for illustrative purposes, the calculation is performed for a coupling as large as \(\eta=0.01\), which is far from the above value~\eqref{eta}.
Nevertheless, the same qualitative features should remain valid at significantly weaker coupling.

\begin{figure}[t]
  \centering
  \includegraphics[width=.6\linewidth]{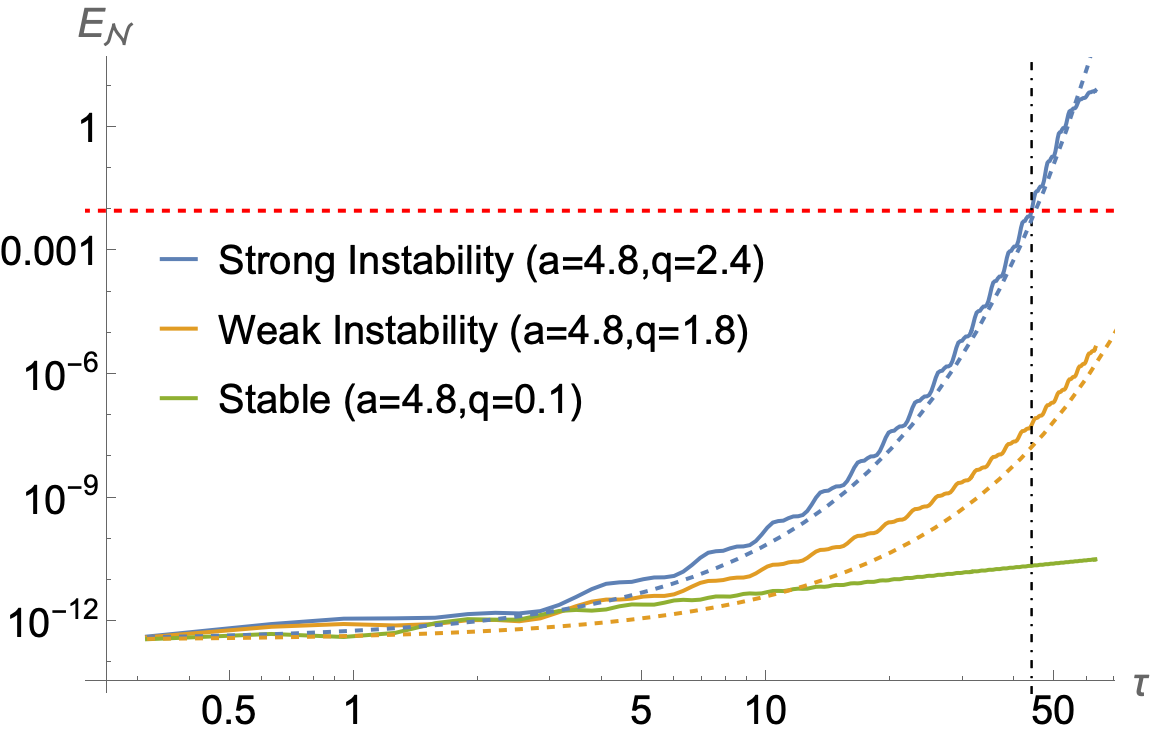}
  \caption{Time evolution of the logarithmic negativity \(E_{\mathcal N}(\tau)\) with coupling \(\eta=10^{-12}\).
  Blue \((a=4.8,\,q=2.4)\), orange \((a=4.8,\,q=1.8)\), and green \((a=4.8,\,q=0.1)\) indicate strongly unstable, weakly unstable, and stable parameters, respectively.
  Dashed lines depict the approximate analytic formula in Eq.\,\eqref{eq:EN-approx-eta} for the unstable cases. No dashed line is shown for the stable trajectory because the Floquet exponent is purely real (\(\mu_I=0\)).
  The blue trajectory reaches the detection threshold \(E_{\mathcal N}=0.01\) (horizontal red dashed line) at \(\tau\simeq 44\) (vertical dot-dashed line), which corresponds to $t \approx 0.044~\mathrm{s}$ for a modulation frequency of $\omega = 1~\mathrm{kHz}$ for instance.  
  }
  \label{fig:timeevo}
\end{figure}

Fig.~\ref{fig:timeevo} presents the time evolution of the logarithmic negativity \(E_{\mathcal N}\) for different instability strengths. We choose three representative parameter points (blue: strongly unstable; orange: weakly unstable; green: stable) as shown in Fig.~\ref{fig:strutt}.
In the unstable cases, \(E_{\mathcal N}\) grows exponentially, and a stronger instability leads to a faster increase. In contrast, in the stable case, \(E_{\mathcal N}\) rises almost linearly in time and is much slower than in the unstable cases.
These results clearly indicate that the magnitude of \(\mu_I\) controls the generation rate of entanglement.

We now focus on the strongly unstable case, shown by the blue line in Fig.~\ref{fig:timeevo}.
We adopt \(E_{\mathcal N} \ge 0.01\) as an experimentally detectable threshold for entanglement~\cite{Palomaki2013}.
Under a weak gravitational coupling of \(\eta = 10^{-12}\), the trajectory crosses this threshold at \(\tau \simeq 44\).
For \(\omega = 1~\mathrm{kHz}\), this corresponds to a physical time of \(t \simeq 0.044\,\mathrm{s}\).
This is much shorter than the typical time scale (\(\sim 3\,\mathrm{s}\)) in 
the original gravity-induced entanglement proposals~\cite{Bose2017,Marletto2017}
and in simply released-mass scenarios~\cite{Krisnanda2020}, indicating that operating in unstable regions can dramatically accelerate entanglement generation.

We also provide an approximate analytic formula for the generation of \(E_{\mathcal N}\) shown by the dashed curves in Fig~\ref{fig:timeevo},
\begin{equation}
E_{\mathcal N}(\tau)\;\simeq\;
\frac{3}{8}\eta
\,e^{2\mu_I\tau}.
\label{eq:EN-approx-eta}
\end{equation}

In our previous work~\cite{Fujita2023}, we have rigorously derived $E_{\mathcal N}\simeq (8/3)\eta\,\exp[2\tau]$ for a pair of inverted oscillators. Since the exponential growth rate is $\mu_I$ in present setup, we replace the exponent $2\tau$ with $2\mu_I \tau$. A reasonably good agreement with the exact result is observed in Fig.~\ref{fig:timeevo}.
The agreement becomes particularly good at late times, where the exponential growth dominates, while the deviation at early times can be attributed to the fact that the Mathieu oscillators are not yet fully governed by the exponential factor and the $\mathcal{O}(1)$ periodic function remain significant.
Also, the plot with the larger $q$ value (blue line) shows a better fit within the range shown in Fig.~\ref{fig:timeevo}. This is because
its Mathieu exponent is bigger and the exponential growth dominates in shorter time scale.

The entanglement generation in our system can be intuitively understood as follows. Since the Hamiltonian includes the gravitational potential $V_{\rm grav}(\hat x_1,\hat x_2)$, the joint state of the two oscillators acquire a position-dependent relative phase. 
Schematically, an initially separable state $\phi(x_1)\phi(x_2)$ evolves into an entangled state due to such a phase $e^{-iV_{\rm grav}(x_1, x_2)t}\phi(x_1)\phi(x_2)$.
As the wavefunction spreads over the position space, the phase difference increases, and in the unstable regions it grows exponentially. This gravity-induced phase shift is the main source of entanglement and this picture is consistent with Eq.~\eqref{eq:EN-approx-eta} which includes only the contribution from exponential spreading.


\section{Environmental decoherence and linear damping}
\label{sec:damping_noise}

We have so far considered an idealized situation without environmental effects.
In this section we incorporate a linear damping and a stochastic force 
into the equations of motion to account for dissipation and decoherence, and we
evaluate gravity-induced entanglement.

We extend Eq.~(\ref{eq:coupled-second}) to include a damping rate $\gamma$ and random
forces $\xi_i(\tau)$:
\begin{equation}
\frac{d^2\hat X_i}{d\tau^2}
+\gamma\frac{d\hat X_i}{d\tau}
+\bigl[a-2q\cos(2\tau)\bigr]\hat X_i
+ \eta\,\hat X_j
=\xi_i(\tau)\,,
\label{eq:HL12}
\end{equation}
where $i,j=1,2$ and $i\neq j$.
The term $\xi_i(\tau)$ represents the additive force noise. In a realistic setting, other noise sources, such as frequency noise (e.g., fluctuations in the parameters $a$ or $q$), would also be present~\cite{Weiss2021}. However, we assume that the decoherence is dominated by the additive force noise.

The random forces are modeled as mutually independent white noises,
\begin{equation}
\langle \xi_i(\tau)\rangle=0,\qquad
\big\langle \xi_i(\tau)\xi_j(\tau')\big\rangle
=\mu\,\delta_{ij}\,\delta(\tau-\tau')\,.
\label{eq:noise}
\end{equation}
The noise strength $\mu$ is related to the damping rate $\gamma$ through a temperature $T$ as  $\mu=\frac{2k_\text{B} T}{\hbar \omega} \gamma$ by the fluctuation-dissipation theorem~\cite{Giovanetti2001, Landau1958}.
This relation is valid for a high temperature regime $k_\text{B} T  \gg \hbar \omega$. 

Using the change of variables (\ref{pm variable}) and defining
$\xi_\pm=(\xi_1\pm\xi_2)/\sqrt2$, the above equations decouple as
\begin{equation}
\frac{d^2\hat X_\pm}{d\tau^2}
+\gamma\frac{d\hat X_\pm}{d\tau}
+\Bigl[(a\pm \eta)-2q\cos(2\tau)\Bigr]\hat X_\pm
=\xi_\pm(\tau).
\label{eq:pm}
\end{equation}
To remove the first derivative, we introduce a new variable
$\hat U_\pm(\tau)=e^{\gamma\tau/2}\hat X_\pm(\tau)$.
Then $\hat U_\pm$ satisfy the Mathieu equation with external forces,
\begin{equation}
\frac{d^2\hat U_\pm}{d\tau^2}
+\left[\left(a\pm \eta-\frac{\gamma^2}{4}\right)-2q\cos(2\tau)\right]\hat U_\pm
=e^{\gamma\tau/2}\,\xi_\pm(\tau).
\label{eq:Ueq}
\end{equation}
It implies that $\tilde C_\pm(\tau)\equiv C(a\pm \eta-\gamma^2/4,\,q,\,\tau)$ and
$\tilde S_\pm(\tau)\equiv S(a\pm \eta-\gamma^2/4,\,q,\,\tau)$ denote the even/odd
solutions of the homogeneous part to the Mathieu euqation for $\hat U_\pm$.
The retarded Green function for the original variable $\hat X_\pm$ reads
\begin{equation}
G_\pm(\tau,s)
=\Theta(\tau-s)\,
\frac{e^{-\gamma(\tau-s)/2}}{\tilde W_\pm}\,
\Bigl[\tilde C_\pm(\tau)\,\tilde S_\pm(s)-\tilde S_\pm(\tau)\,\tilde C_\pm(s)\Bigr]\,,
\label{eq:Gpm}
\end{equation}
where $\Theta(\tau)$ is the Heaviside step function and $\tilde W_\pm=\tilde C_\pm(0)\,\frac{d\tilde S_\pm(0)}{d\tau}$ is the constant Wronskian.
Then we find the general solution
\begin{align}
\hat X_\pm(\tau)
&= e^{-\gamma\tau/2}\!\big[\hat A_\pm\,\tilde C_\pm(\tau)+\hat B_\pm\,\tilde S_\pm(\tau)\big]
   - \int_{0}^{\tau}\!ds\,G_\pm(\tau,s)\,\xi_\pm(s)                                      \\
&\equiv \hat{D}_\pm(\tau)+N_\pm(\tau)\,,
\end{align}
where we have defined $\hat D_\pm$ and $N_\pm$ as the homogeneous and the inhomogeneous (noise) solution, respectively, for later use.

Having obtained the general solution, we now proceed to compute the covariance matrix.
Since the full calculation is rather lengthy, we illustrate the procedure by explicitly evaluating
the $(1,1)$ component as a representative example.
It is given by
\begin{align}
V_{11}(\tau)=\langle \hat X_1(\tau)^2\rangle
= \frac12\!\left[\langle \hat X_+^2\rangle+\langle \hat X_-^2\rangle
                 +\langle \hat X_+\hat X_-\rangle+\langle \hat X_-\hat X_+\rangle\right]\,.
\end{align}
The auto-correlations are computed as
\begin{align}
\langle \hat X_\pm^2\rangle
&=\langle \hat D_\pm^2\rangle+\langle N_\pm^2\rangle\,,
   \\[2pt]
\langle \hat D_\pm(\tau)^2\rangle
&= e^{-\gamma\tau}\!\Big[\tilde C_\pm(\tau)^2\langle \hat A_\pm^2\rangle
  + \tilde S_\pm(\tau)^2\langle \hat B_\pm^2\rangle
  + \tilde C_\pm(\tau)\tilde S_\pm(\tau)\langle \hat A_\pm\hat B_\pm+\hat B_\pm\hat A_\pm\rangle\Big]\,,
  \\
\langle N_\pm(\tau)^2\rangle
&= \mu \int_{0}^{\tau}\!ds\, G_\pm(\tau,s)^2\,,
\end{align}
where the cross terms such as $\langle \hat D_\pm N_\pm\rangle$ vanish because $\langle \xi_\pm(s)\rangle=0$. 
We also use the white-noise correlations 
$\langle \xi_\pm(s)\xi_\pm(s')\rangle=\mu\delta(s-s')$ and $\langle \xi_\pm(s)\xi_\mp(s')\rangle=0$ which follow from Eq.~\eqref{eq:noise}.
The cross-correlations are 
\begin{align}
\langle \hat X_\pm(\tau)\hat X_\mp(\tau)\rangle
= e^{-\gamma\tau}\!\Big[\tilde C_\pm\tilde C_\mp\,\langle \hat A_\pm\hat A_\mp\rangle
 + \tilde C_\pm\tilde S_\mp\,\langle \hat A_\pm\hat B_\mp\rangle
 + \tilde S_\pm\tilde C_\mp\,\langle \hat B_\pm\hat A_\mp\rangle
 + \tilde S_\pm\tilde S_\mp\,\langle \hat B_\pm\hat B_\mp\rangle\Big],
\end{align}
Putting them altogether, we obtain the $(1,1)$ entry of the covariance matrix as 
\begin{align}
V_{11}(\tau)
&= \frac{e^{-\gamma\tau}}{2}\Big\{
\tilde C_+^2\langle \hat A_+^2\rangle+\tilde S_+^2\langle \hat B_+^2\rangle
+\tilde C_+\tilde S_+\,\langle \hat A_+\hat B_++\hat B_+\hat A_+\rangle \notag\\
&\qquad\qquad\quad
+\tilde C_-^2\langle \hat A_-^2\rangle+\tilde S_-^2\langle \hat B_-^2\rangle
+\tilde C_-\tilde S_-\,\langle \hat A_-\hat B_-+\hat B_-\hat A_-\rangle \notag\\
&\qquad\qquad\quad
+\tilde C_+\tilde C_-\,\langle \hat A_+\hat A_-\rangle
+\tilde C_+\tilde S_-\,\langle \hat A_+\hat B_-\rangle
+\tilde S_+\tilde C_-\,\langle \hat B_+\hat A_-\rangle
+\tilde S_+\tilde S_-\,\langle \hat B_+\hat B_-\rangle\Big\} \notag\\
&\quad + \frac{\mu}{2}\int_{0}^{\tau}\!ds\,\big[G_+(\tau,s)^2+G_-(\tau,s)^2\big].\label{41}
\end{align}
All other components of $V(\tau)$ can be calculated in the same way.

We then impose the separable Gaussian initial state $V(0)=\tfrac12 I_4$ together with the canonical
commutators. These conditions fix the undetermined operator coefficients as
\begin{equation}
\langle \hat A_\pm^2\rangle=\frac{1}{2\,\tilde C_\pm(0)^2},\qquad
\langle \hat B_\pm^2\rangle=\frac{4+\gamma^2}{8\,\left(\tfrac{d \tilde S_\pm(0)}{d\tau}\right)^2}
,\qquad
\langle \hat A_\pm\hat B_\pm\rangle=-\langle \hat B_\pm\hat A_\pm\rangle\frac{2i\pm\gamma}{4\,\tilde C_\pm(0)\tfrac{d \tilde S_\pm(0)}{d\tau}},
\label{42}
\end{equation}
and
\begin{equation}
\langle \hat A_+\hat A_-\rangle=\langle \hat A_+\hat B_-\rangle
=\langle \hat B_+\hat A_-\rangle=\langle \hat B_+\hat B_-\rangle=0.
\label{43}
\end{equation}
In the limit \(\gamma \to 0\), the expressions reduce to the previously obtained non-dissipative results Eqs.~\eqref{eq:AB-moments} and \eqref{14}. 
Substituting Eqs.~\eqref{42} and \eqref{43} into Eq.~\eqref{41}, and introducing the normalized functions 
\(\bar{C}_{\pm}(\tau)\equiv \tilde{C}_{\pm}(\tau)/\tilde{C}_{\pm}(0)\) and \(\bar{S}_{\pm}(\tau)\equiv \tilde{S}_{\pm}(\tau)/\frac{d\tilde{S}_{\pm}(0)}{d\tau}\) we obtain the explicit $(1,1)$ entry
\begin{align}
V_{11}(\tau)
&=\frac{e^{-\gamma\tau}}{16}\Big\{4\,\!\big[\bar C_+(\tau)^2+\bar C_-(\tau)^2\big]
+(4+\gamma^2)\!\big[\bar S_+(\tau)^2+\bar S_-(\tau)^2\big]
+4\gamma\,\!\big[\bar C_+(\tau)\bar S_+(\tau)+\bar C_-(\tau)\bar S_-(\tau)\big]\Big\}\notag\\
&\quad+\frac{\mu}{2}\int_0^\tau\!du\,e^{-\gamma(\tau-u)}\!
\left\{\big[\bar C_-(\tau)\bar S_-(u)-\bar S_-(\tau)\bar C_-(u)\big]^2
+\big[\bar C_+(\tau)\bar S_+(u)-\bar S_+(\tau)\bar C_+(u)\big]^2\right\}.
\end{align}

All the remaining entries of $V$ are obtained analogously. Based on the full covariance matrix, we evaluate the logarithmic negativity \(E_{\mathcal N}\) to quantify the generated entanglement.

\begin{figure}[t]
  \centering
  \includegraphics[width=.6\linewidth]{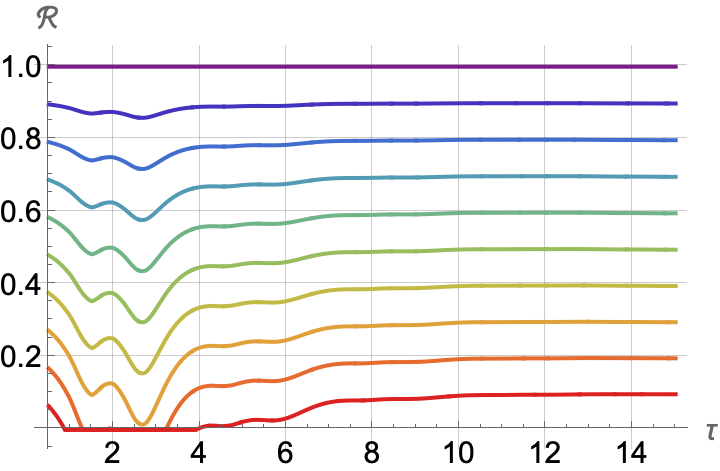}
    \caption{We plot the ratio $\mathcal{R}$ defined in Eq.~\eqref{Ratio R} as a function of $\tau$ to illustrates how environmental random forces suppress the logarithmic negativity $E_{\mathcal N}$. We fix the gravitational coupling at $\eta=10^{-5}$ and use the Mathieu parameters $(a,q)=(4.8,\,2.4)$. The top horizontal line corresponds to $\mu=0$.  From top to bottom, the random-noise strength increases as $\mu = 0,\,0.1\,\eta,\,0.2\,\eta,\,\ldots,\,0.9\,\eta$, in steps of $0.1\,\eta$.  
  As $\mu/\eta$ increases, the entanglement is progressively suppressed, 
  and the late-time plateau value of $E_{\mathcal N}$ asymptotically approaches $\mathcal{R} = 1-\mu/\eta$.
  }
  \label{mu_rainbow}
\end{figure}
We first examine the case where only the random noise force is taken into account as the environmental effect, neglecting dissipation ($\gamma=0$).
Fig.~\ref{mu_rainbow} presents the reduction of the entanglement due to the random noise force for $\eta=10^{-5}$ and $\gamma=0$.
We define the ratio between the logarithmic negativity with and without the random noise force as
\begin{equation}
\mathcal{R}(\tau) = 
\frac{E_{\mathcal N}\big|_{\mu}}{E_{\mathcal N}\big|_{\mu=0}}\,.    
\label{Ratio R}
\end{equation}
Alhough a transient dip is observed at early times, the ratio $\mathcal{R}$ approaches a plateau with a value of $1-\mu/\eta$ in the long time limit.
This implies that under the influence of random forces, 
the logarithmic negativity behaves as
\begin{equation}
    E_{\mathcal N}(\tau\gg 1)\propto \eta-\mu\,.
    \label{noise}
\end{equation}
Hence, the noise strength $\mu$ should be kept smaller than $\eta$ in order to realize gravity-induced entanglement.

\begin{figure}[t]
  \centering
  \includegraphics[width=.6\linewidth]{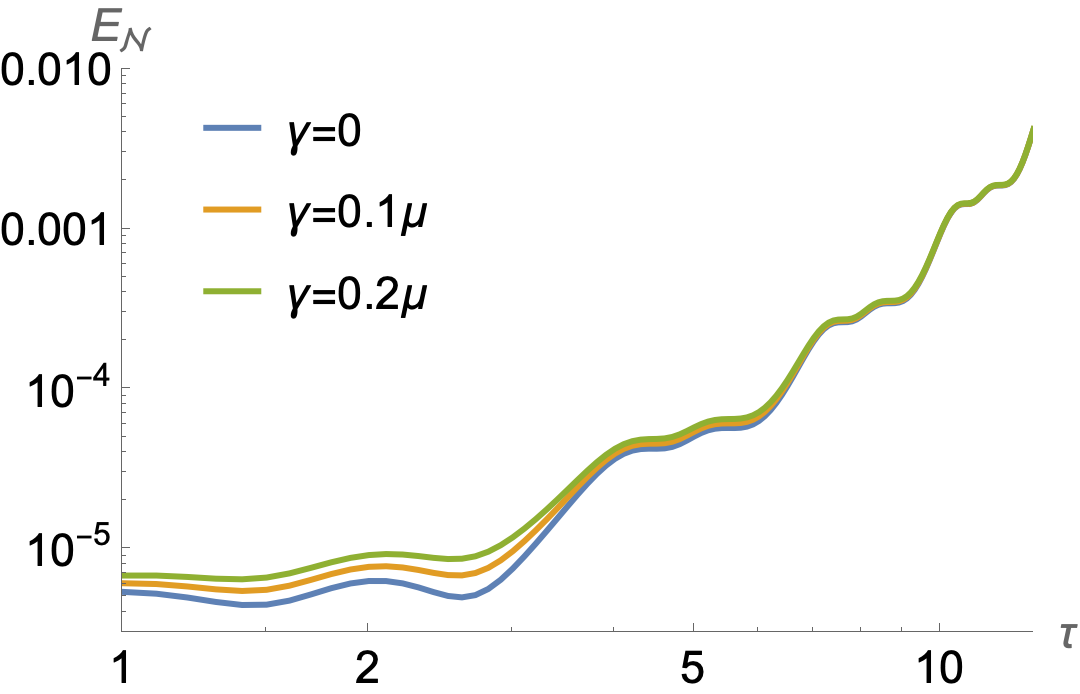}
  \caption{Effect of linear damping on the logarithmic negativity $E_{\mathcal N}(\tau)$.  
The linear damping rate is set to $\gamma=0$ (blue), $0.1\,\mu$ (orange), and $0.2\,\mu$ (green). 
These values of $\gamma$ are consistent with the fluctuation-dissipation theorem and used here merely as illustrative examples.
Other parameters are fixed at $\eta=10^{-5}$, $\mu=5\times10^{-6}$, and $(a,q)=(4.8,\,2.4)$.  
In this parameter regime, the quantum uncertainty relation is satisfied, $\det[V(\tau)] \ge 1/16$,  
and the green curve with $\gamma=0.2\,\mu$ corresponds to the saturation value of this bound.  
The damping effect slightly enhances $E_{\mathcal N}$ at early times, but the enhancement becomes negligible when  
$E_{\mathcal N}$ reaches the detectable level of $10^{-2}$.}
  \label{gammma}
\end{figure}

Finally, we incorporate both the random forces and the damping effect. Fig.~\ref{gammma} shows
$E_{\mathcal N}(\tau)$ for several damping rates $\gamma$ at fixed $\eta$ and $\mu$.
Interestingly, we observe that as the linear damping increases, 
the logarithmic negativity $E_{\mathcal N}(\tau)$ is enhanced.
One might be tempted to think that introducing a large damping could further enhance the entanglement.
However, a significant enhancement cannot be achieved in this way.  
According to the quantum uncertainty relation, the determinant of the covariance matrix must satisfy $\det[V] \ge 1/16$.  
If the damping is increased too much,  $\det[V]$ falls below this bound.  
Thus, the damping rate has a fundamental upper limit. The green curve in Fig.~\ref{gammma} corresponds to nearly this saturation value,  
but the enhancement of $E_{\mathcal N}$ is visible only at early times;  
by the time $E_{\mathcal N}$ grows to the detectable level, the difference has already become negligible.  
We conclude that, in our system, damping does not prevent the generation of entanglement,  
but neither does it significantly assist it.


\section{Conclusion}
\label{sec:conclusion}

We analyzed the generation of gravity-induced entanglement between two parametrically driven oscillators. 
Based on the analytic solution, we constructed the covariance matrix and evaluated the logarithmic negativity $E_{\mathcal N}$.
When the imaginary part of the Floquet exponent $\mu_I$ is positive, 
the system enters an unstable regime where the logarithmic negativity grows exponentially. 
In the idealized case without environmental effects, and for parameter choices deep in the unstable regime but still within the range 
where the effective spring constant remains non-negative, 
the exponential growth of the logarithmic negativity
is approximately given by \eqref{eq:EN-approx-eta}.
For instance, entanglement reaches a detectable level 
\(E_{\mathcal N}\approx0.01\) around \(\tau\simeq44\), for instance,
corresponding to a physical time of \(0.044~\mathrm{s}\) with \(1~\mathrm{kHz}\) modulation.

We found that random force noise suppresses the entanglement generation.
The degree of suppression is characterized by the ratio \eqref{noise} 
between the noisy and noiseless cases (with and without $\mu$). 
Here $\mu$ represents the strength of the white noise force. 
This ratio provides a practical requirement on how low the noise level should be 
relative to the gravitational coupling $\eta$. 
When the noise strength $\mu$ is smaller than the coupling $\eta$, 
the logarithmic negativity still exponentially grows, allowing gravity-induced entanglement 
to develop even in the presence of random noise.
We also include a linear damping rate $\gamma$  and evaluate its impact on gravity-induced entanglement under the restriction of the uncertainty relation. Interestingly, $\gamma$ slightly increases the logarithmic negativity. However, the slight increase in the logarithmic negativity due to $\gamma$ becomes negligible after the exponential growth of entanglement. Therefore, $\gamma$ does not hinder entanglement generation, nor does it effectively enhance it in our system.

One practical route is to realize a Mathieu-type drive via the optical-spring effect, which provides a tunable time-dependent stiffness in a cavity???mechanical setup~\cite{Rossi2025,Ma2020}. 
A complementary route is to apply a small periodic modulation to the effective length of the pendulum system~\cite{Hartwig2023}.
Another important challenge is to address the competition between gravitational coupling and environmental random noise. In our Mathieu-coupled setup, just as in previous proposals, we find that entanglement cannot be generated  unless the noise strength $\mu$ is kept below the gravitational coupling $\eta$. To make an experiment feasible in the near future, we will need strategies that go beyond a simple $\eta$???$\mu$ comparison and effectively suppress decoherence. From this perspective, a key objective for future work is to identify configurations in which gravity-induced entanglement is efficiently generated even in noisy environments.


\begin{acknowledgements}
We would like to thank Yuta Michimura for fruitful discussions. This work was supported in part by the Japan Society for the Promotion of Science (JSPS) KAKENHI, Grants No. JP23K03424 (T.F.), JP23K13103 (A.M.), 24KJ1233 (Y.K.) 
and by Grant-in-Aid for JSPS Fellows (Y.K.).
\end{acknowledgements}

\bibliography{ref.bib}

\end{document}